# Perceptual Requirements for
# World-Locked Rendering in AR and VR

Phillip Guan
Reality Labs Research, Meta
United States of America
philguan@meta.com

Eric Penner
Reality Labs Research, Meta
United States of America
epenner@meta.com

Joel Hegland
Reality Labs Research, Meta
United States of America
hegland@meta.com

Benjamin Letham
Meta
United States of America
bletham@meta.com

Douglas Lanman
Reality Labs Research, Meta
United States of America
douglas.lanman@meta.com

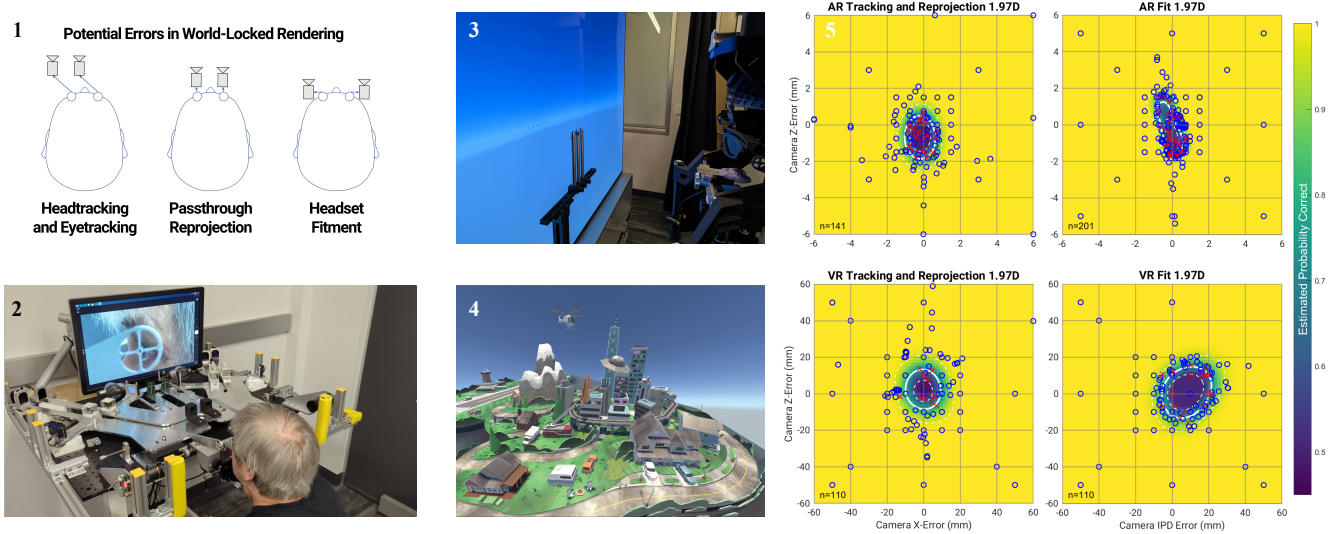

Figure 1: We present a psychophysical study to determine render camera placement requirements to achieve world-locked rendering (WLR) in augmented and virtual reality (AR and VR). (1) Potential sources of error leading to render camera displacement from user eye position in head-mounted displays. (2) A bite bar calibration and measurement system used to precisely position users' head and eyes. (3) Calibrated bite bars are used in a rotating chin rest to enable highly precise, perspective-correct rendering. Here three optical posts serve as real-world references to judge the stability of virtual spheres rendered above them for AR WLR requirements. (4) The scene used to study WLR in VR. (5) Study results for one observer showing acceptable camera displacement errors to achieve WLR in AR and VR (note 10x difference in scale). Blue and red points show trials where camera displacement error is correctly or incorrectly identified; white contours represent WLR error detection thresholds (75% performance).

## ABSTRACT

Stereoscopic, head-tracked display systems can show users realistic, head-locked virtual objects and environments. However, discrepancies between the rendering pipeline and physical viewing conditions can lead to perceived instability in the rendered content resulting in reduced immersion and, potentially, visually-induced motion sickness. Precise requirements to achieve perceptually stable world-locked rendering (WLR) are unknown due to the challenge of constructing a wide field of view, distortion-free display with highly accurate head and eye tracking. We present a system capable of rendering virtual objects over real-world references without perceivable drift under such constraints. This platform is used to study acceptable errors in render camera position for WLR in augmented and virtual reality scenarios, where we find an order of magnitude difference in perceptual sensitivity. We conclude with an analytic model which examines changes to apparent depth and visual direction in response to camera displacement errors.

studies in HCI.**

## 1 INTRODUCTION

Head-mounted displays (HMDs) can track a user's movement and
position to present perspective-correct binocular views of virtual
content. When headset tracking, rendering, and presentation pipelines
are accurate HMD users will perceive stabilized and world-locked
virtual objects. In practice, these HMD subsystems are imperfect
and errors in head and eye tracking, rendering latency, imprecise
distortion correction, and poor headset fit while wearing HMDs and
their physically correct counterparts. These differences can lead to
instability in rendered content and incomplete world-locked rendering (WLR). Such artifacts are particularly noticeable in augmented
reality (AR) headsets because the real world provides a reference to
highlight these errors. However, virtual reality (VR) and mixed reality (MR) headsets are also affected by WLR errors which can lead to
perceived local or global instability in VR as well as misregistration
of the virtual and real world in MR.

Several analyses that examine geometric errors from perspective-
incorrect rendering have been introduced [Holloway 1997; Hwang
and Peli 2019; Rolland et al. 2004; Wann et al. 1995; Woods et al.
1993]. However, triangulation-based solutions to depth do not
always exist. Monocular occlusion can lead to da Vinci stereopsis [Nakayama and Shimojo 1990] and certain errors in stereoscopic
displays can result in non-intersecting skew rays—corresponding
features without a geometric solution to depth [Held and Banks
2008]. The visual system can also interpret visual cues based on
context, for example, the surface slant of artwork (or displays) can
be estimated from its frame and used to partially compensate for
distortions from off-axis viewing [Banks et al. 2014; Vishwanath
et al. 2005]. Thus a comprehensive understanding of requirements
for WLR must incorporate aspects of human perception in addition
to geometric modeling.

Only a handful of studies explicitly consider the perceptual consequences of inconsistent render camera and user viewing positions
(Section 2), and within this work WLR accuracy is only considered
for stationary observers. WLR errors for non-moving observers
results in static virtual to world misalignment in AR or static shape
distortion in VR. With user motion static errors become dynamic
and manifest as unintended motion in the virtual content relative
to the world in AR and shape or scene instability in VR. Importantly, these dynamic errors are more easily detectable than static
ones [Mckee and Nakayama 1984]. Additionally, user head and body movements introduce visual-vestibular cue integration, which affect how inaccurate visual cues are interpreted [Butler et al. 2010;
Cuturi and MacNeilage 2014; Fetsch et al. 2009; Fulvio et al. 2021;
MacNeilage et al. 2012]. These studies highlight the importance of
studying WLR with active observers (i.e., allowing for some degree
of natural eye and head movements).

There are two major complications when designing a WLR user
study with active observers. First, no off-the-shelf HMDs render
or track accurately enough to achieve complete WLR for a moving
observer. Second, AR and VR headsets or large cave autonomous
virtual environments (CAVE) that support active observers are also
affected by significant display artifacts that introduce additional,
non-generalizable geometric errors to stimuli used in experiments.
We address these challenges in studying WLR requirements for
HMDs with the following contributions:

- We introduce the first system capable of complete world-
  locked rendering for active observers in AR by introducing
  hardware to generate highly-accurate head and eye models
  used for perspective-correct rendering in a distortion-free,
  wide field of view stereoscopic display for users performing
  a vestibulo-ocular reflex eye movement.
- We conduct a user study using this hardware to derive perceptual requirements for world-locked rendering in AR and
  VR headsets, providing the first direct AR and VR comparison of such requirements. We investigate the implications of
  different sources of errors and compare the effects of tracking inaccuracy and headset fit discrepancies on WLR. We
  further consider the implications of headset virtual display
  distance and, for VR, the effect of scene content.
- We compare study results to a geometric model and highlight
  visual direction in addition to depth errors from incorrect
  disparity as a potential predictor for human sensitivity to
  WLR errors in VR.

## 2 RELATED WORK

*User Studies on Render Camera Placement.* Pollock et al. [2012]
and Kelly et al. [2013] explicitly evaluate the impacts of incongruent
viewing and rendering positions. Both studies place users away
from the renderer center of projection (CoP) by several feet and
measure the errors in perceived depth, distance, or direction in a
CAVE and find that the perceived errors are smaller than predicted
by geometric models. Ponto et al. [2013] describe a user-in-the-loop
calibration for a VR HMD to improve WLR with optimized rendering parameters. With this procedure they report more accurate
depth judgments and reduced spatial drift from different viewpoints,
but the values obtained for interpupillary distance (IPD) in their
user-calibrated method are, on average, 16 mm larger than measured values suggesting that other errors exist in the rendering
pipeline which are compensated for by this large discrepancy.

A recurring discussion in 1990s HMD research was whether parallax from millimeter-scale movement of the user's CoP during an
eye rotation was perceptually significant [Holloway 1997; Rolland
et al. 2004; Wann et al. 1995]. These small shifts, coined *ocular parallax* by Brewster [1845] over 100 years earlier, have been a topic
of recent interest, and studies have found them to be detectable
under certain conditions [Konrad et al. 2020; Krajancich et al. 2020;

Lee et al. 2015]. Importantly, all existing work on WLR has been in display systems with optical distortions and/or inaccurate head and eye tracking systems. Furthermore, these studies are also conducted with stationary observers. The hardware, software, and user study described in this work address these two concerns and our results provide additional context for when different types and magnitudes of error are important to consider in AR and VR headsets.

*Studies using Headsets.* Research related to perception in HMDs is limited by the capabilities of available hardware so studies have primarily focused on how accurately HMD viewers perceive an intended scene or object rather than perceptual requirements needed to eliminate potential artifacts. El Jamiy and Marsh [2019] and Creem-Rhegar et al. [2023] survey many of these recent studies, which are often related to how accurately depth is perceived. Most studies find that distance to objects is underestimated in HMDs, but, due to limits in tracking accuracy and distortion correction, it is difficult to determine how much this phenomenon can be attributed to shortcomings in hardware [Itoh et al. 2021] or to geometric artifacts introduced from inaccurate rendering and distortion correction. A recent study by Wilmott et al. [2022] on WLR uses a commercially available AR HMD (HoloLens 2). They added 10 Hz noise to world-locked objects and found that users could detect this motion at 1-3 arcmin amplitudes. Although limited by the tracking and rendering systems in the HoloLens 2, these results provide a notable point of comparison for our study results in Section 4.

*Other Cues to Depth.* A significant body of work in vision science has studied which retinal cues the visual system uses to estimate absolute depth and distance. Even in well-controlled settings, distance and depth are misperceived without monocular texture cues (i.e., when using perspective-free stimuli to study depth constancy) [Bradshaw et al. 1996; Cumming et al. 1991; Guan and Banks 2016; Johnston 1991]. Hartle and Wilcox [2022] compare real objects with rendered counterparts and, with the introduction of perspective cues, subjects are able to perceive accurate depth in virtual environments, but only when they view the real objects first. Zhong et al. [2021] go further, and use a multifocal plane display with high dynamic range to create rendered images that are indistinguishable from reality under specific viewing conditions. Their use of the multifocal plane display is important because accommodation and defocus cues can be more important than disparity in depth perception [Burge and Geisler 2011; Held et al. 2012; Watt et al. 2005; Zannoli et al. 2016]. While our study investigates the importance of accurate pictorial cues for depth perception, fixed-focus displays without accurate blur and accommodation cues could be insufficient to achieve complete WLR under all conditions.

*Acceptable Video See-Through.* Research in video see-through and video passthrough have found that users can adapt to significant mismatches (55-125 mm) between the original camera position and the user's eyes [Biocca and Rolland 1998; Lee et al. 2013]. Even larger errors are possible with perceptual adaptation over the course of several days [Lee and Park 2020]. Our work aims to determine the smallest errors leading to inaccurate depth perception and does not consider potential adaptation of the visual system to suprathreshold errors. Notably, adaptation may improve user experience in VR and MR HMDs, but cannot resolve WLR conflicts in AR.

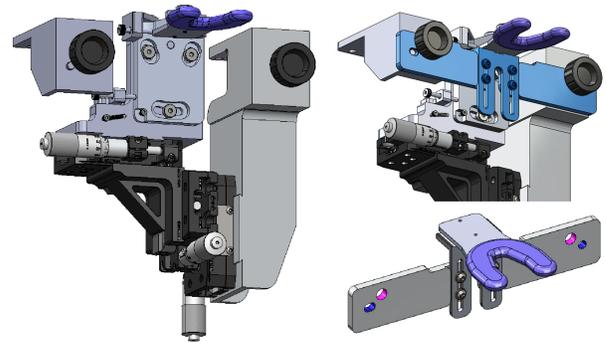

**Figure 2:** *Five Degree-of-Freedom Bite Bar Adjustment Stage. (Left) Precision adjustment stage with 3D translation and 2D rotation used in our bite bar calibration and measure platform (Figure 1.2). (Right) After calibration a modular baseplate (in blue) is attached to the bite bar and the assembly is transferred to the rotating chin rest in Figure 1.3.*

## 3 WORLD-LOCKED RENDERING TESTBED

Here, we describe a WLR-capable system that can support active observers with a distortion-free, wide-FOV display and highly accurate head and eye tracking. The shutter-glasses display and tracking schemes are inspired by Guan et al. [2022] with additional improvements required to achieve WLR described below.

### 3.1 Wide-FOV Display

An 88" LG OLED TV (OLED88ZXPUA), placed 50.7 cm from the user's eyes was used to achieve a 125° x 94° FOV. A pair of active shutter glasses are synchronized to the display with a photodiode to enable temporally interlaced stereoscopic presentation to the eyes at 120 Hz (60 Hz per eye). Images for the left and right eyes are rendered asynchronously which mitigate potential depth artifacts from asynchronous stereo viewing [Hoffman et al. 2011].

### 3.2 Head and Eye Tracking

User head movement is restricted in a rotating chin rest (Figure 1.3) to facilitate accurate head movement and eye tracking. A custom bite bar (Figure 2) is built for each user that places their eyes 9.3 cm away from a rotary encoder and the user's head movements are restricted to yaw rotation only, resulting in head movement similar to shaking "no." Users are instructed to fixate on a known location while rotating their heads which induces a vestibulo-ocular reflex (VOR) that causes user's eyes to move by counter-rotating against the head to stabilize the fixation point. The position of the eyes during VOR can be estimated precisely with an accurate model of the user's head and eyes combined with their head rotation and position of the fixation point in 3D space. The customized bite bar is generated on the bite bar calibration platform described in Section 3.3. A spherical eye model, further described by Krajancich et al. [2020], with a 7.8 mm radius from the center of rotation (CoR) to center of projection (CoP) and foveal fixation using the visual axis rather than the optical axis is used to model the eye rotation for all users during their VOR.

### 3.3 Bite Bar Calibration Platform

A novel bite bar calibration and measurement platform is used to ensure accurate CoP tracking during VOR (Figures 1.2 and 2). This hardware positions each user's head on the bite bar so that their corneal apexes are 9.3 cm away from the chin rest's CoR when the bite bar is used in the chin rest. Head position is precisely manipulated using a high-resolution, five degree-of-freedom rotation and translation stage (x, y, z, roll, and yaw) to position both corneal apexes to a target eye relief and height in combination with a camera-based alignment system (see supplementary videos). A monocular boresight alignment task is used to measure each eye's lateral position on the bite bar after both eyes are positioned at the target nominal height and eye relief. The bite bar is then secured to a baseplate with adhesive (Loctite HY 4070) and transferred to the rotating chin rest. The baseplate serves as a common frame of reference between both systems which allows the head measurements made in the calibration platform to be used for rendering and tracking in the rotating chin rest system.

### 3.4 Accurate Latency Prediction

To achieve WLR a perspective-correct image must be rendered from the user's eye position when the images are seen which requires forward prediction to account for image rendering and display presentation time. We apply a second-order Savitzky-Golay filter on the rotary encoder signal to predict the user's eye positions after any rendering and display latency. More accurate prediction can be achieved when the prediction time is small, so we first optimized the software stack to reduce total latency to 16.67 ms as reported by an nVidia Reflex latency and latency measurement tool[1]. We next measured motion-to-photon latency with the wide-FOV display from Section 3.1 at 20-26 ms. This value was obtained using an oscilloscope to determine the time between changes in display luminance triggered by changes in the rotary encoder signal. The additional 10 ms of latency in this measurement can most likely be attributed to one frame of processing on the OLED panel and 1-2ms of USB latency from the rotary encoder. We independently hand-tuned the time constant for forward prediction and found that 26 ms best achieved WLR as described in Section 3.5.

### 3.5 End-to-End WLR Evaluation

Geometric analysis (Section 5) shows render camera placement errors are most consequential for content rendered away from a near display plane (both in dioptric units). A representative scenario in an HMD under these conditions is hands-based interaction with a varifocal display [Padmanaban et al. 2017] where the user's hands are closer than the nearest focal distance of the headset. A real-world reference will highlight errors in rendering for WLR so we created an AR scene with the display 50.7 cm (1.97 diopters) and three optical posts 38.3 cm (2.61 diopters) away from the user's eyes. The posts are placed at the edge of the Zone of Comfort [Shibata et al. 2011] to maximize potential rendering errors while following user experience best practices. A four millimeter threaded insert is added to each optical post and a five millimeter virtual sphere is rendered above each insert (Figure 1.3). Users evaluate the accuracy of WLR by fixating on the center threaded post while rotating their head. Any errors in the WLR system will cause the virtual spheres to drift relative to each threaded insert [Azuma and Bishop 1994], and users determine whether or not the spheres remain stationary over the posts. In this evaluation the most discerning observers could detect a small amount of drift in the spheres. Krajancich et al. [2020] allowed users to make small, fine-tuned adjustments to align virtual and physical content in their AR study, and complete WLR could be achieved when users were allowed to make small initial adjustments (1-2 mm) to align the optical posts to the virtual spheres with their heads at the nominal 0° rotation viewing position.

## 4 WLR USER STUDIES

Render camera displacement errors for stationary observers result in misalignment of virtual content to the real world (AR) and shape distortions (AR and VR). The same displacement errors will introduce unintended motion and dynamic distortions when the user moves as incorrect perspective geometry results in inconsistent shape and depth across viewpoints. The visual system is more sensitive to these dynamic artifacts and we use the WLR-capable system from Section 3 to run a psychophysical user study with active observers to determine render camera positioning requirements for WLR. Additional study details and raw data are presented in the supplementary materials.

### 4.1 User Study Stimulus and Task

Detectable render camera displacement was identified using a two-interval forced choice task (2IFC). In one interval render cameras were co-located with the CoP and in the other interval render cameras were displaced from the CoP. Participants were asked to fixate on a specified gaze target during a horizontal VOR eye movement, but they were not instructed to make their VOR movements with any particular magnitude or frequency. They controlled the speed, duration, and magnitude of their head rotation. The virtual scene varied depending on experimental settings, but within each condition, scenes were identical in each interval with the exception of render camera displacement. The two intervals were randomly ordered and subjects were asked to identify which interval appeared more world stable and, for VR scenes, less distorted. Acceptable camera placement was defined by the errors that could be accurately identified 75% of the time. Auditory feedback was provided after each trial to indicate whether participants correctly identified the trial with render camera displacement. The initial eight trials in each experiment were set to suprathreshold values and the auditory feedback allowed researchers to verify that participants understood the study protocol. Auditory feedback does not influence outcomes or reproducibility, but makes participants happier so it was left in beyond the training trials [Bach and Schäfer 2016].

Eleven subjects participated in the user study (three male, eight female, ages 20-46), all subjects were screened for 40 arcseconds stereoacuity with a Randot test and 20/20 visual acuity with a Snellen eye chart. Participants requiring corrective eyeglasses were excluded from the study in order to minimize eye relief with our shutter glasses (contact lenses were allowed). All subjects participated in the four main conditions (Figure 3) and a subset participated in conditions examining the impacts of display distance and VR content (Figures 4 and 5). All study protocols were IRB approved.

---

[1]https://developer.nvidia.com/nvidia-latency-display-analysis-tool

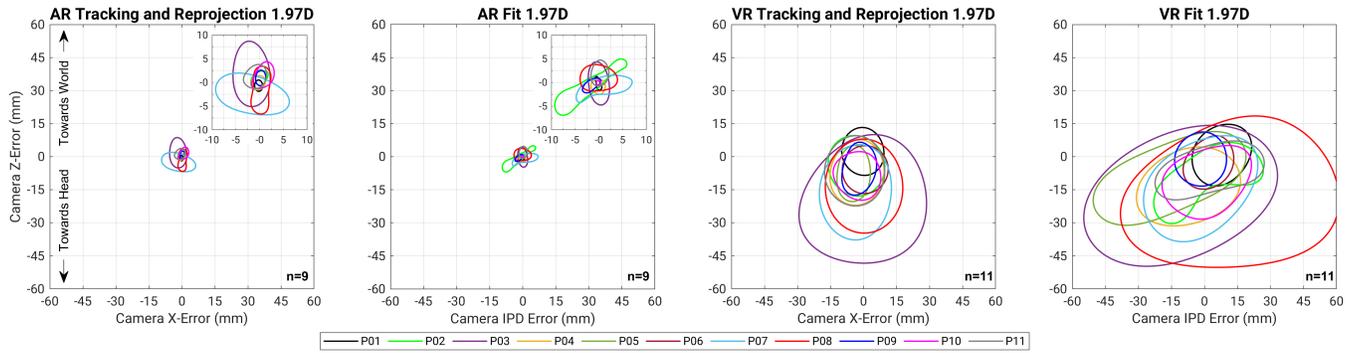

**Figure 3:** *WLR Requirements for Near Display.* 75% correct detection thresholds for camera displacement relative to both eye's CoP leading to disruptions in WLR for a 1.97 D (51 cm) display for all subjects. The display distance is selected based on criteria outlined in Section 3.5. For all conditions the ordinate represents a relative longitudinal offset between the user's eyes and render cameras (i.e., eye relief). Positive values are towards the display and negative values are towards the user. For tracking conditions Camera X-Error represents a leftwards or rightwards shift. For fit conditions Camera IPD error represents the total error in camera baseline relative to the user's true IPD (e.g., a -10 mm IPD error is a 5 mm nasal shift from each eye). All errors are in the head frame of reference.

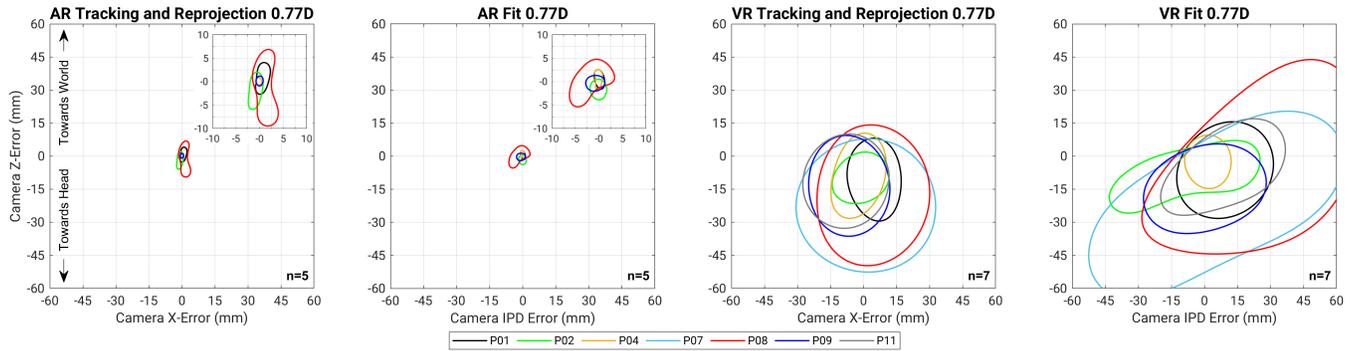

**Figure 4:** *WLR Requirements for Far Display.* 75% correct detection thresholds for camera displacement relative to both eyes' CoP leading to disruptions in WLR for a 0.77 D (1.3 m) display for all subjects. Data was collected for the same subjects as the near display condition (Figure 3), but a smaller subset of subjects participated in this study due to time constraints and availability for followup data collection sessions.

## 4.2 Modeling and Threshold Extraction

We used AEPsych [Owen et al. 2021] to collect data and learn a model for the user response (probability the error was detected) as a function of lateral and longitudinal camera displacement. For response curve modeling, camera displacement was represented in polar coordinates and the model enforced monotonicity in detection probability with respect to displacement magnitude; see the supplementary materials for more details.

## 4.3 WLR Requirements for AR and VR

In Experiment 1 render camera placement requirements are derived for AR and VR viewing conditions. The AR scene is described in Section 3.5 and users were instructed to fixate on the middle threaded screw. For VR, the optical posts were removed and a different scene was shown (Figure 1.4). In the VR scene the fixation point is located on a cartoon person at the center of the image, and the fixation point is positioned at the center of the display.

Render camera displacements are simulated independently in each eye, and the relative displacement of the cameras has distinct perceptual consequences. In the lateral dimension opposite sign errors represent an IPD error and same sign errors are equivalent to a head tracking error to the left or right. In the longitudinal axis same sign errors represent an error in eye relief, and opposite sign errors place the render cameras at different depths, slightly magnifying one eye's image relative to the other. In the vertical dimension same sign errors represent an error in the user's height and opposite sign errors introduce a global vertical disparity shift. The testbed hardware does not allow vertical head movement, so errors in the vertical axis are not considered in this study. Two combinations of error relevant to HMD usage were considered in the study. The first represents head tracking or camera reprojection errors that displace the user's head and are simulated with same sign errors in the lateral and longitudinal axes (x-error and z-error respectively). The second represents headset fit errors which occur when the assumed IPD and eye relief values used in rendering are not accurate (opposite sign x-error and same sign z-error).

User sensitivity to tracking and fit errors are shown for AR and VR conditions in Figure 3. Two subjects, P05 and P06, were not able to reliably identify camera displacements at the ±15 mm error limits used in AEPsych adaptive data collection in the AR conditions. Their modeled detection thresholds extend beyond errors actually presented during data collection so these results are excluded (Figures S6 and S7 in supplementary materials). Using the area within each participant's thresholds to define acceptable camera displacement, AR tolerances for tracking and fit errors are 31.4 mm² ± 36.5 mm² and 21.1 mm² ± 18.2 mm². The same thresholds are more than an order of magnitude larger in VR, 7.6 cm² ± 7.0 cm² and 17.0 cm² ± 15.8 cm², respectively. A paired t-test across the nine participants who successfully completed both AR and VR conditions shows a statistical difference between the thresholds (AR: 26.3 mm²±28.5 mm², VR: 13.4 cm²±13.7 cm², p=.0007). Sensitivity to errors are not statistically significant when comparing tracking and fit conditions in AR (p=.35), but are significant in VR (p=.02). Considering the ordinate is the same for tracking and fit conditions, this indicates that lateral tracking errors disrupt WLR more than IPD errors; a surprising finding further explored in Section 5.

Another notable asymmetry between the AR and VR conditions is an apparent bias towards negative eye relief in VR. We compute the centroid of each subject's acceptable camera displacement errors, and in AR there are no significant biases away from zero error, but in VR eye relief errors in both tracking and fit conditions are significantly biased away from zero (tracking: -8.5 cm²±6.2 cm², p=.001; fit:-9.4 cm²±6.5 cm² p=.0007). Taken together, these results indicate that real world references in AR appear to ground render camera positions errors for WLR; errors are centered around zero and thresholds are equivalent between simulated fit and tracking errors. Once physical references are removed in VR, tracking and fit errors are not equivalent, and eye relief errors towards the user's head are less detectable than errors away from the user's head.

### 4.4 Virtual Display Distance

The same camera errors will often lead to larger error for a near display distance [Rolland et al. 2004]. The 1.97 D (51 cm) viewing distance used in Section 4.3 is much closer than display distances on commercially available HMDs which are in the 0.5 D-1.3 D (2 m to 77 cm) range. We repeat the same protocol as before, and move both the physical posts and render plane to 0.77 D (1.3 m) to determine the impacts of virtual display distance on WLR requirements in a configuration more applicable to existing devices. For the AR condition the virtual spheres and optical posts are again 0.64 D in front of the display. However, the FOV reduction of the display at the farther viewing distance forces a trade off between maintaining the same depth range (in diopters) or seeing the entire scene. When the virtual environment is enlarged to maintain depth, a significant portion of the scene is cropped at the larger scale and relative distortions between objects at different depths are no longer visible. We elected to show the same scene without increasing its scale (the prominent features in the scene near the fixation point extend approximately ±30 cm away from the display at both viewing distances). Five participants repeated the AR conditions and seven repeated the VR conditions due to subject availability (Figure 4). The average area for acceptable error at the two viewing distances was compared using a paired t-test yielding no significant difference in sensitivity at the near display distance (16.2 mm² ± 6.9 mm²) and far distance (19.2 mm²±21.8 mm²) for AR (p=0.61). There was a significant difference for VR at the near distance (11.4 cm²±12.6 cm²) compared to the far distance (20.5 cm²±17.9 cm², p=0.01). However, it is unclear if this change can be attributed to display distance or differences in scene depth (in diopters).

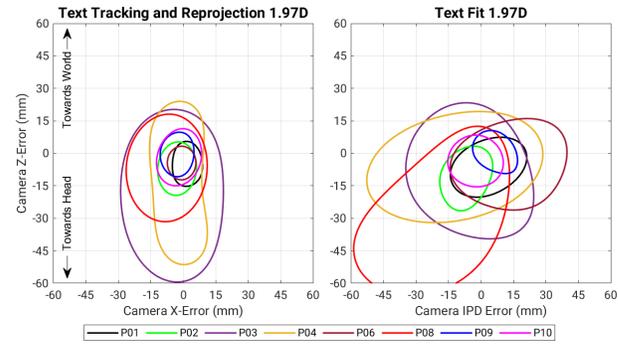

**Figure 5: *Render Camera Displacement Thresholds for Text in VR.* Results are shown for detectable camera displacement in millimeters relative to user CoP positions at the near (1.97 D, 50.7 cm) display distance, but while viewing a plane of slanted text instead of the city scene used for data collection in Figure 3. (Left) Results for simulated head tracking and MR reprojection errors. (Right) Results for simulated headset fitment errors.**

### 4.5 VR Scene Content

Previous work [Guan et al. 2022] found stronger sensitivity to optical distortions in VR while viewing text compared to a more irregular and less familiar 3D scene. We conduct this same comparison for our final study. However, content located at the render display plane is invariant to errors in render camera position [Rolland et al. 2004] which means in a world-locked plane of text at the display plane will always be rendered correctly in our testbed regardless of differences in render and viewing geometry. To account for this we added a 12.8° top-back slant to the text to introduce potential errors from incorrect perspective. This angle was selected to match the slant of the empirical vertical horopter at 1.97 D [Aizenman et al. 2022; Cooper et al. 2011]. We again compare acceptable camera error in the two scenes with a paired t-test and find no significant difference (city: 12.7 cm²±14.7 cm²; text 13.3 cm²±12.5 cm², p=.79).

## 5 ANALYTIC MODELING

Next, we implement an analytic model similar to others identified in Section 2. This model uses binocular perspective projection to identify geometric errors arising from inconsistent render and viewing positions for points within the horizontal meridian defined by the eyes and gaze point. The model works in three stages. First, the viewer's head position, orientation, and gaze are established and the eye model from Krajancich et al. [2020] is used to establish each eye's CoP location. Render cameras are placed at each CoP and the render plane is set to the simulated virtual display distance. Ground-truth disparities and visual direction are then computed for all points in the scene. In the second pass, render cameras are displaced relative to each eye's CoP and points throughout the horizontal meridian are again projected onto the camera projection planes. These inaccurate 2D projections for the left and right eyes are used to calculate erroneous disparity and visual direction information when viewed from the actual CoP positions. In the last stage, ground truth disparities and visual directions are compared to these erroneous values to compute geometric errors for depth and visual direction.

In this model the origin is represented by the original location of the user's cyclopean eye when the head is not rotated, and the angular plots shown in Figures 6 and 7 do not account for small changes in the cyclopean eye's position and rotation during head rotation. The plots also include a 0.6 D range around the display plane representing the Zone of Comfort and, in the disparity plot, Panum's Area is shown representing the region where the visual system can fuse binocular images. Conditions for fusion vary across a number of parameters so this model is approximate. The criteria for fusion was 20 arcmin at fixation [Palmer 1961] with a linear increase to 60 arcmin at 6° eccentricity [Hampton and Kertesz 1983; Mitchell 1966; Palmer 1961]. After six degrees fusion limits increase by 7% per degree of eccentricity [Crone and Leuridan 1973; Mitchell 1966].

*Disparity and Visual Direction.* WLR in AR requires accurate alignment in depth and visual direction, and this is reflected in Section 4.3 with users demonstrating equal sensitivity to IPD errors and lateral displacements in AR. In VR results indicate that lateral tracking errors are more perceivable than IPD errors. This result is unintuitive based on geometric modeling because IPD errors lead to significantly larger distortions compared to head tracking errors.

The columns in Figure 6 show triangulation errors in world space, and disparity and visual direction errors in angular and dioptric space when render and view geometries are mismatched. The top row shows a simulation for -12 mm camera IPD error and the bottom row simulates a -12 mm translation error. The world view reconstruction for IPD error in the top left panel shows significant geometric error. However, our user study results indicate that the errors in the bottom simulation from a translation error are more perceivable (computed areas from Figures 3 and 4). Additional insight into this unexpected result can be found in supplemental videos of the same simulations over the course of a ±20° VOR head and eye movement. In these videos the depth distortions from IPD error cause significant shifts in the reconstructed blue squares compared to the intended gray squares defined by the render geometry. However, because the simulated IPD error is symmetric in both eyes, the visual direction of the blue squares is relatively unaffected. Conversely, the blue squares drift from side to side when simulating tracking error and sensitivity to dynamic errors in visual direction could explain why tracking errors are more detectable than IPD errors. One potential explanation for this phenomenon is that depth cues from motion parallax, which are relatively unaffected by IPD errors, help counteract the inaccurate depth cues from binocular disparity when there is IPD error, but motion from changes in visual direction lead to global scene instability.

*Comparison to Existing Hardware.* Next we consider how thresholds from Section 4 compare to capabilities of existing HMDs. Tracking performance and fit statistics are not readily available, so we instead turn to previous research. The most analogous study to ours was conducted by Wilmott et al. [2022] who added 10Hz sinusoidal motion in a random direction to world-locked objects in HoloLens 2. They found that this motion was detectable at 1-3 arcmin amplitude for seated observers viewing a 10° floating cube. This protocol is not directly comparable to ours, as their observers were stationary and were tasked to detect when world-locked cubes appeared to move. The continuous 10Hz motion added to the cubes is likely more detectable than the errors we simulated, but the tracking and display systems of the HoloLens 2 also make errors in WLR more difficult to detect (less precise head tracking, lower display resolution, waveguide color non-uniformity). Our data show -1.5 mm of error in eye relief and IPD are detectable for about half our observers in AR. We examine the impacts of these errors in the geometric model (Figure 7) and see that such errors induce a 4 arcmin disparity error at the fixation point and a total change in visual direction of 3.6 arcmin. Similarly, -1.5 mm eye relief with a -1.5 mm tracking error leads to 3.3 arcmin of visual direction change and disparity errors ranging from 0 to 0.9 arcmin at the fixation point. The magnitude of these errors are well above thresholds for stereo and visual acuity, but our AR conditions also included 0.64 D of vergence-accommodation conflict. Although these results are similar to Wilmott et al. [2022], further investigation is necessary to understand the implications of differences in study stimuli.

## 6 DISCUSSION

We present a WLR-capable system achieved with the introduction of novel bite bar fabrication and calibration hardware. Using this WLR testbed, we ran the first user study to directly compare the perceptual requirements for WLR render camera placement for AR and VR viewing environments. The study results are analyzed in the context of a perspective-projection based model where apparent visual direction is identified as an important consideration for WLR in VR. We conclude with closing thoughts and potential implications for HMD design.

*Implications of Constrained User Movement.* Our WLR testbed restricts user head and eye movement to a horizontal VOR in order to estimate user head and eye position with high accuracy and minimal latency. Enforcing consistent head motion across trials also enables comparison of results across all trials, conditions, and participants by eliminating differences in perceived drift from variations in head and eye movement. While the results from our experiment are specific to horizontal VOR, the detection thresholds detailed in Section 4 are potentially more stringent compared to WLR requirements for other head and eye movements because smooth pursuits and saccades introduce motion blur to non-fixated image features and saccadic suppression. Both phenomena will likely reduce sensitivity to WLR errors identified in our user study conducted with VOR.

*Pupil Swim.* Gaze-contingent distortions from near-eye optics can introduce additional errors to HMD users even if the rendered images are perspective-correct. This distortion, known as pupil swim, arises from changes in the user's eye position relative to HMD optics and can only be addressed by eye-tracked dynamic distortion correction (ETDCC). The effects of pupil swim have been studied by others [Guan et al. 2022; Kuhl et al. 2008; Tong et al. 2019, 2020], and we do not consider the combined effects of pupil swim and render camera placement because the perceptual effects of pupil swim are highly dependent on optical design and do not generalize, unlike the WLR thresholds outlined in our study.

*Ocular Parallax.* HMDs without eye tracking cannot update the camera position to account for gaze-contingent ocular parallax (Section 2). Eye rotations are generally less than 20° [Aizenman et al. 2022], and a typical eye with a 7.8 mm distance from the CoR to CoP at a 20° rotation would translate 2.8 mm. Our user study results indicate that 1-5 mm of error leads to detectable errors in WLR for AR, so it will be necessary to account for ocular parallax to achieve WLR in AR. The same tolerances in VR are 10 mm or larger, so rendering ocular parallax may not be necessary for WLR in VR. However, our user study does render ocular parallax and the magnitude of this parallax is consistent with a user eye, even if error added to the camera position results in incorrect perspective projection. Regardless of the potential significance of ocular parallax in VR, optical distortions in VR resulting from the same eye movements can induce significant optical distortions that are larger than errors from incorrect render camera position [Geng et al. 2018; Robinett and Rolland 1992; Rolland and Hopkins 1993]. If ETDCC is used to address pupil swim then it is trivial to use the tracked eye position for rendering. Eye tracking requirements for ETDCC are on the millimeter scale [Guan et al. 2022], so eye tracking systems with millimeter-level accuracy are required for both AR and VR headsets to achieve perceptually stable virtual content, albeit for potentially different reasons.

*Mixed Reality.* Showing users the video feed from exterior-facing cameras on VR HMDs and rendering virtual content with the video stream has emerged as a middle ground between fully-immersive VR and world-blended AR. Like normal rendering, passthrough video cameras in VR headsets will suffer from WLR errors when the cameras are not positioned at the user's true eye position. MR is fully immersive and does not contain real world references to highlight perspective errors, so we will apply VR thresholds to evaluate WLR in MR. Our user study results identify 15 mm as an approximate displacement threshold for WLR in MR (Figures 3 and 4). The simplest solution for MR passthrough is direct passthrough, which has the added benefit of minimizing latency, but even extremely thin diffractive optics [Maimone and Wang 2020] are unlikely to achieve the 15 mm total thickness required for WLR. Alternatively, novel optical designs can be used to optically place a passthrough camera closer to the user's eyes [Kuo et al. 2023; Takagi et al. 2000] and such designs which should aim to be within 15 mm of the nominal eye location. Finally, reprojection algorithms using novel view synthesis [Chaurasia et al. 2020; Xiao et al. 2022] can also be employed in combination with the above solutions for passthrough. Our user study results indicate that errors behind the eyes are less detectable compared to errors in front of the eyes, and view reprojection algorithms could reproject to a nominal location slightly behind the user's true eye position to minimize the effects of potential reprojection errors in these algorithms.

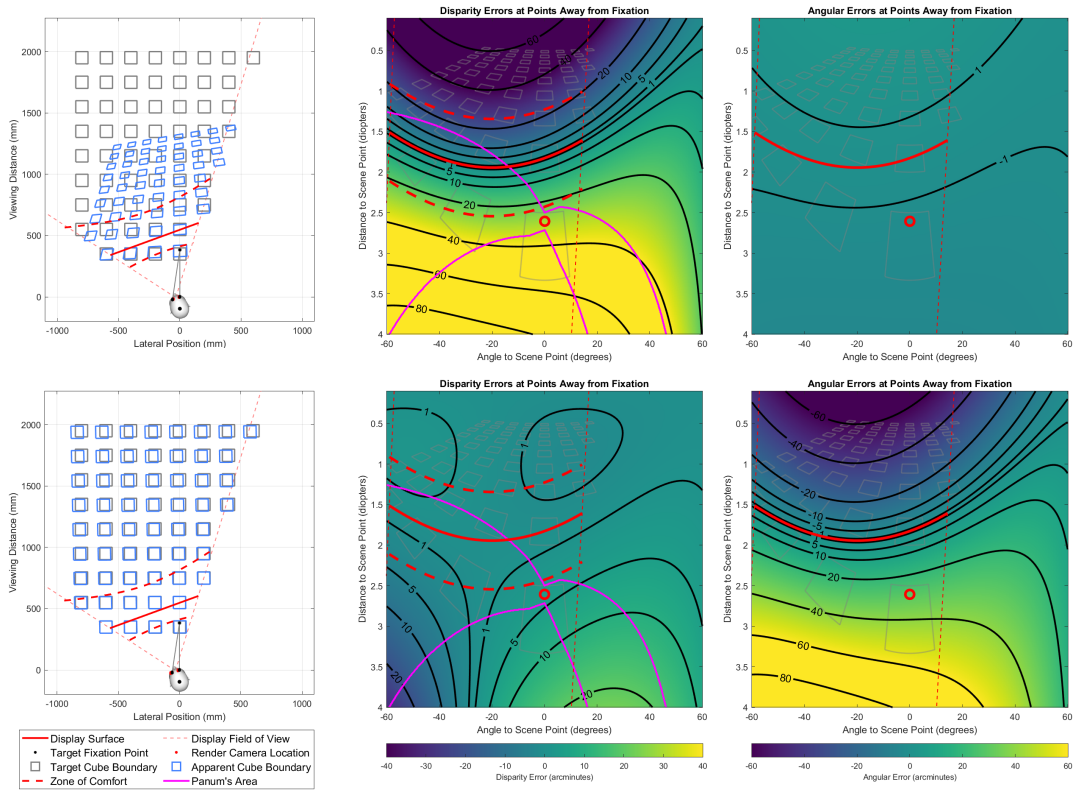

**Figure 6:** *Analytic Model of IPD and Tracking Errors in VR.* **(Top)** A 63 mm IPD observer with a -20° head rotation viewing a scene rendered with 51 mm IPD camera baseline. **(Bottom)** The same user viewing content rendered with a -12 mm displacement instead. **(Left)** A reconstruction of the scene in world space showing depth and misalignment errors in the scene. **(Center)** Disparity errors for points away from fixation compared to ground truth. **(Right)** Errors in visual direction. Our user study shows that the impacts of disparity errors are less apparent to observers compared to changes in visual direction over the duration of a ±20° VOR head movement.

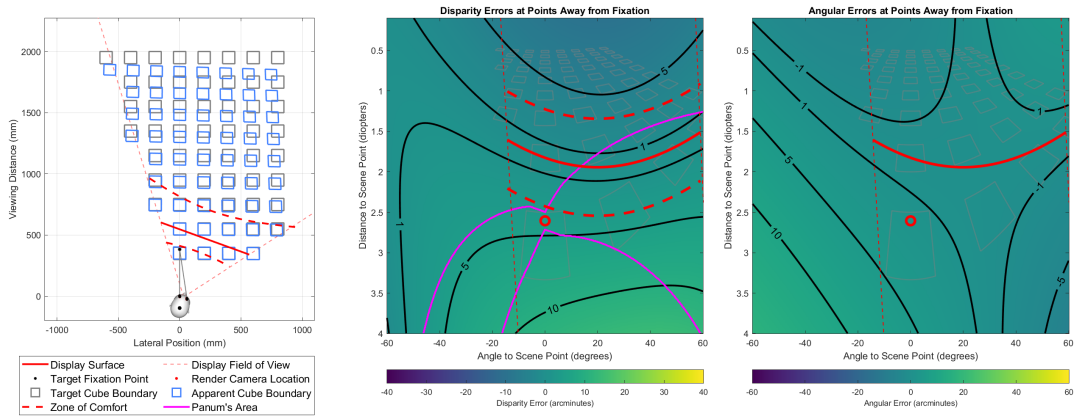

**Figure 7:** *Analytic Model of Errors in AR.* Geometric consequences of a 63 mm IPD observer with a 20° head rotation viewing a scene rendered with -1.5 mm error in eye relief and camera separation. **(Left)** A reconstruction of the scene in world space showing depth and misalignment errors in the scene. **(Center)** Disparity errors for points away from fixation compared to ground truth. **(Right)** Errors in visual direction. Modeling results show similar thresholds found by Wilmott et al. [2022] using a HoloLens 2. We find similar thresholds in the 1-3 arcmin range when applying render camera errors from our results, however, our study protocol and hardware are not directly comparable and similarities in thresholds need additional verification.

# Perceptual Requirements for World-Locked Rendering in AR and VR
## Supplementary Material

ANONYMOUS AUTHOR(S)
SUBMISSION ID: 126

# 1 ADDITIONAL USER STUDY DETAILS

## 1.1 Response Surface Modeling

We used models to estimate the response surface in the user study, probability of detection as a function of camera displacement. As in other recent work [Gardner et al. 2015; Guan et al. 2022; Owen et al. 2021], we used a Gaussian process (GP) as a non-parametric model to flexibly estimate the response surface without specifying a particular form *a priori*. We represent camera displacement in polar coordinates as $(r, \theta)$, and denote the probability-of-correct-response function as $p(r, \theta)$. We model $p(r, \theta) = \Phi(f(r, \theta))$ where $\Phi$ is the Gaussian cumulative distribution function and $f(r, \theta)$ is a latent response function that has a GP prior. Given data $\mathcal{D}$ collected in the user study, the GP prior on $f$ induces a normal posterior distribution $f(r, \theta)|\mathcal{D} \sim \mathcal{N}(\mu, \sigma^2)$, whose mean and variance can be computed in closed form.

For the WLR user study, we know that detection probability will be monotonic in the magnitude of the displacement, $r$, since larger errors will always be easier to detect. We incorporate that knowledge into the model by enforcing monotonicity in $f$ with respect to $r$. We do so by projecting the posterior distribution in an adaptation of the GP projection method of [Lin and Dunson 2014]. Our method differs from that of [Lin and Dunson 2014] by doing projection of quantiles rather than projection at the sample level, to improve computational efficiency. Specifically, when computing the posterior at $f(r, \theta)$, we construct a uniform sequence of $m + 1$ points $\tilde{r}_i = \frac{ri}{m}$, $i = 0, \ldots, m$, that range from 0 up to $r$; we used $m = 20$ in practice. Denote the posterior at this sequence as $f(\tilde{r}_i, \theta)|\mathcal{D} \sim \mathcal{N}(\mu_i, \sigma_i^2)$. We apply a monotonic projection first to the mean at $(r, \theta)$ by computing $\hat{\mu} = \max_{i=0,\ldots,m} \mu_i$. We similarly project a lower and upper quantile as $l = \max_{i=0,\ldots,m} \mu_i - 2\sigma_i$ and $u = \max_{i=0,\ldots,m} \mu_i + 2\sigma_i$ respectively. This construction ensures that $\hat{\mu}$, $l$, and $u$ will all be monotonic with respect to $r$, for any fixed $\theta$. Finally, we use these projected moments as the posterior for $f(r, \theta)$: specifically, $f(r, \theta) \sim \mathcal{N}\left(\hat{\mu}, \left(\frac{u-l}{4}\right)^2\right)$. If $f$ was already monotonic with respect to $r$ then this projection will have no effect, and so with sufficient data the projected posterior distribution will converge to the true, monotonic posterior.

## 1.2 User Study Protocol

We elected to employ a two-interval forced choice (2IFC) protocol shown in Figure 1 to identify the thresholds in our study instead of a faster, but more subjective yes/no task to reduce variability across subjects. Data was collected in two hours per session with subjects coming in for two to four sessions depending on availability. Each condition was run with AEPsych using 25 to 32 initialization points after which adapative sampling was used to collect data up to 110 trials. Subjects collected data in two hour blocks on separate days, and subjects participated in two to five sessions depending on availability. In general subjects were able to complete between three to six different conditions in each two hour block of time. In general we aimed to collect at least 110 trials for each condition, but collected additional trials when time permitted. For adaptive sampling with AEPsych the search limits were set to ±15mm in AR conditions and ±60cm for IPD and ±100mm for eye relief in the VR conditions. For the AR conditions a different maximum value was used to model the data using the method describe in Section 1.1 based on the each subject's sensitivity. This is because using the maximum limits for every subject during model fitting is inefficient for subjects with data clustered near smaller values (e.g the modeled area for P01 in Figure 2, especially for the near display distance, is mostly empty beyond 3mm). The limits used for each participant are represented by the limits shown in the individual raw data plots In Figures 2 through 12.

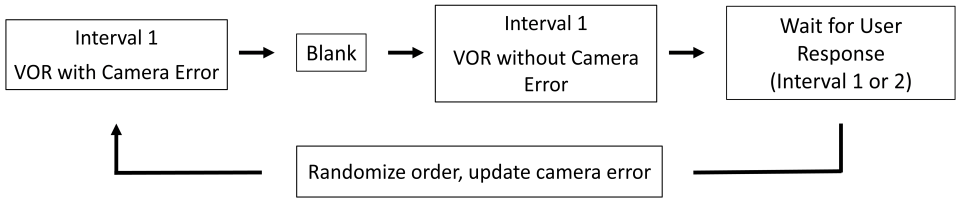

Fig. 1. *Two-interval forced choice procedure:* In the 2IFC task participants compare pupil swim during VOR to a distortion-free reference in every trial compared to a yes/no procedure where subjects compare potential distortion during VOR to a mental model of a distortion-free reference. 2IFC reduces variance across subjects because every trial is compared against the same reference, but increase data collection time.

## 1.3  AR User Study Data

*1.3.1  AR Tracking and Reprojection; Fit; and Viewing Distance.* Here we show data for each participant for all AR conditions.

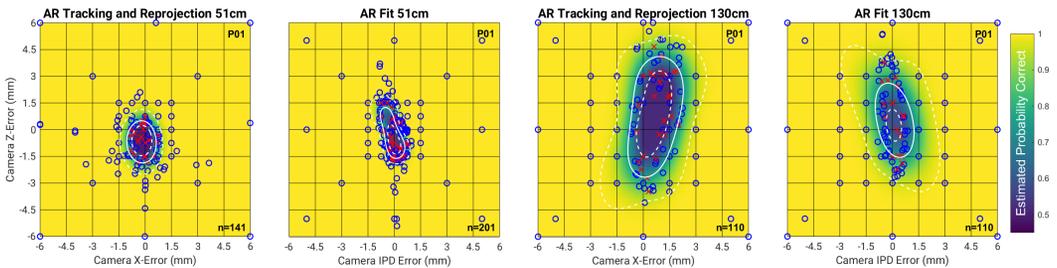

Fig. 2. P01 Raw data for near and far display distances for AR conditions.

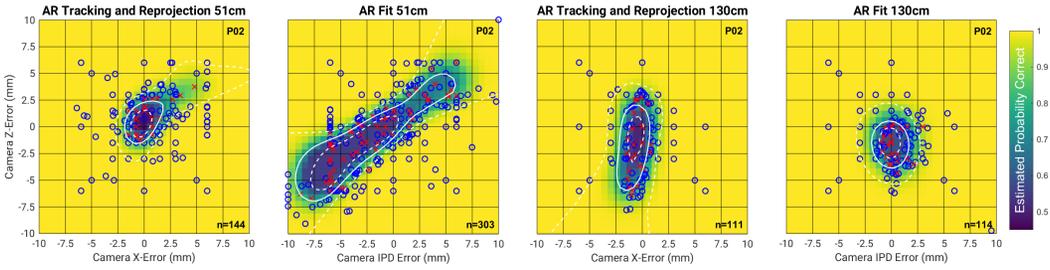

Fig. 3. P02 Raw data for near and far display distances for AR conditions.

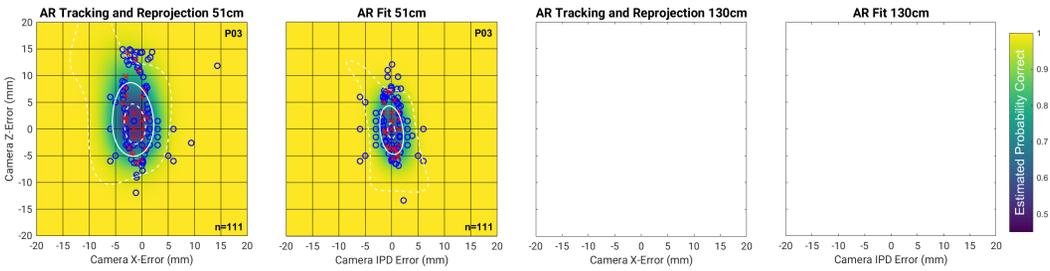

Fig. 4. P03 Raw data for near and far display distances for AR conditions.

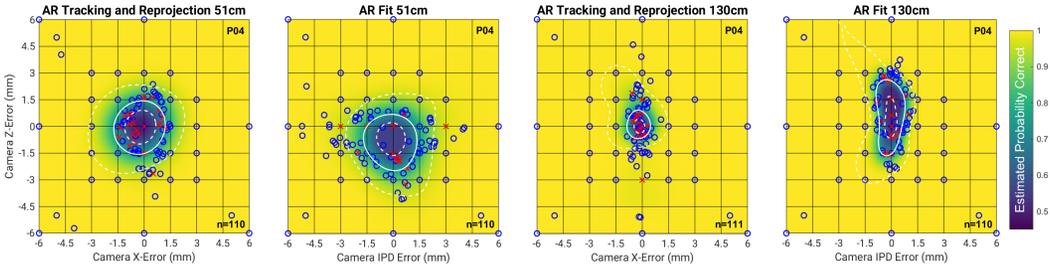

Fig. 5. P04 Raw data for near and far display distances for AR conditions.

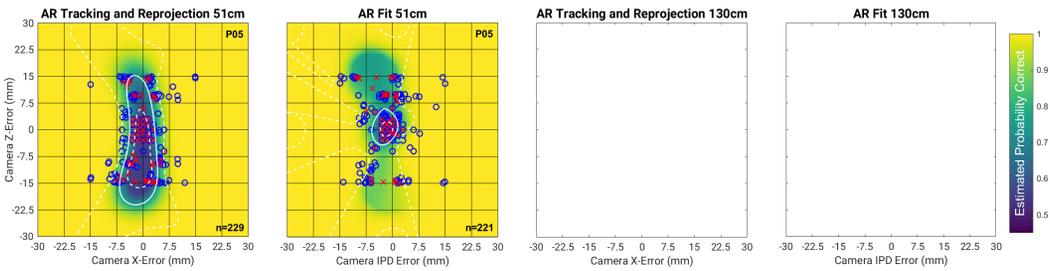

Fig. 6. P05 Raw data for near and far display distances for AR conditions. This subject was not able to reliably complete the AR task at the limits supplied to AEPsych (thresholds are not surrounded by blue circles indicating consistently correct responses) so their AR thresholds are excluded from Figure 3 in the main paper.

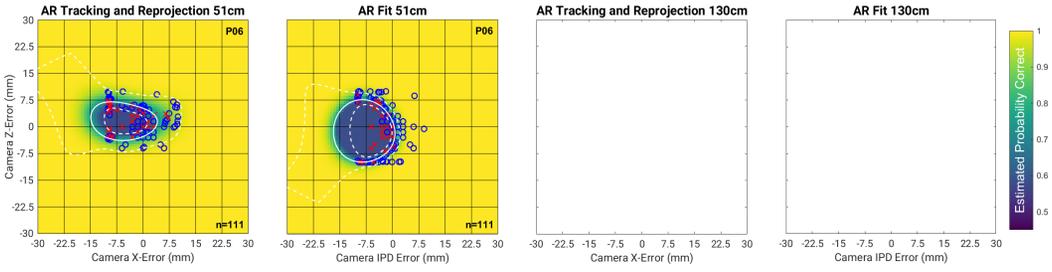

Fig. 7. P06 Raw data for near and far display distances for AR conditions. This subject was not able to reliably complete the AR task at the limits supplied to AEPsych (thresholds are not surrounded by blue circles indicating consistently correct responses) so their AR thresholds are excluded from Figure 3 in the main paper.

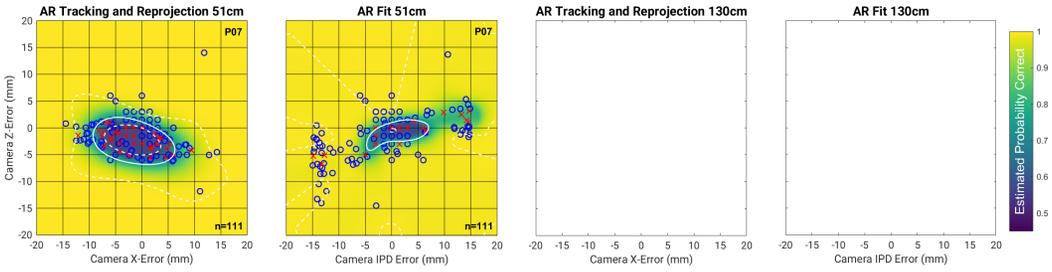

Fig. 8. P07 Raw data for near and far display distances for AR conditions.

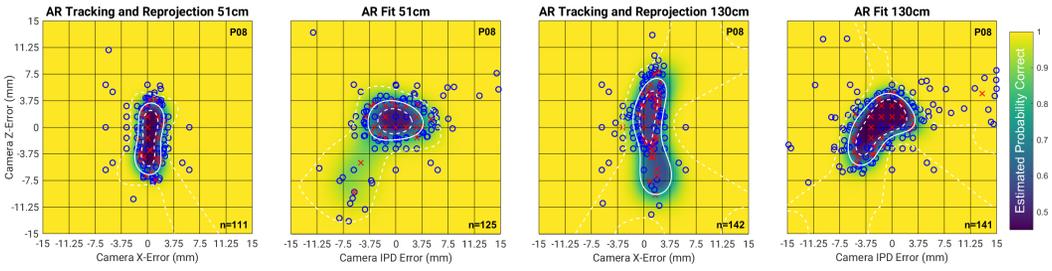

Fig. 9. P08 Raw data for near and far display distances for AR conditions.

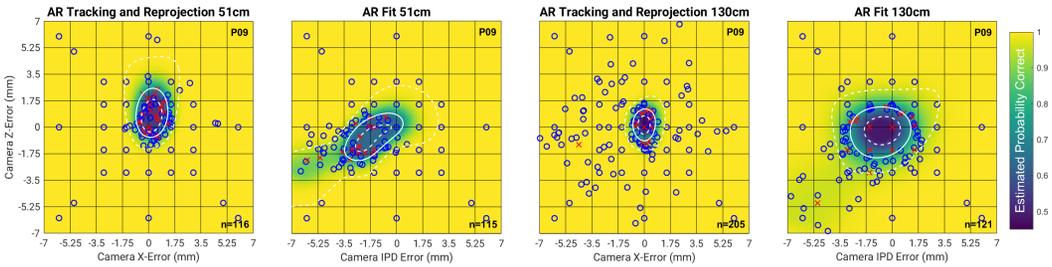

Fig. 10. P09 Raw data for near and far display distances for AR conditions.

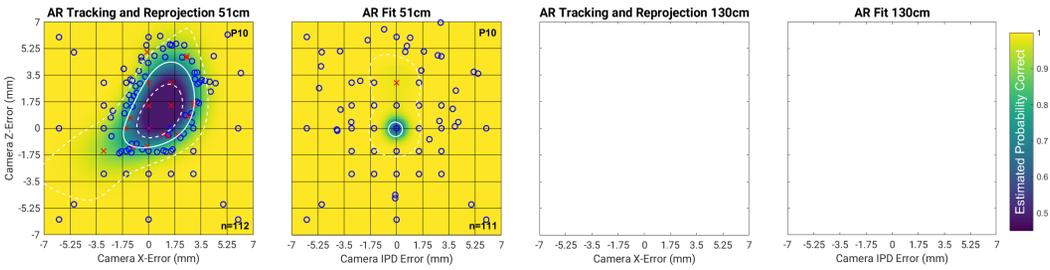

Fig. 11. P10 Raw data for near and far display distances for AR conditions. Note: a lapse trial by this participant and an unusually accurate performance near zero resulted in too many suprathreshold trials for this participant in the AR fit condition. See Section 1.3.2 for more details.

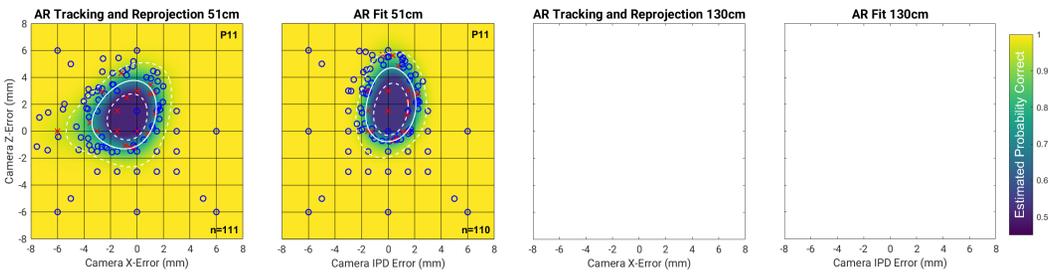

Fig. 12. P11 Raw data for near and far display distances for AR conditions.

*1.3.2 Adaptive Sampling.* Figure 13 shows the trials run for P10. In this instance a lapse trial at a very large value, combined with only a single missed trial near (0,0) resulted in the model sampling large values rather than zero. During data collection we elected to use the built-in Gaussian Process classification model from AEPsych and this model does not have strong priors which means that inefficient sampling can occur in edge cases such as this where a lapse trial and very few missed trials result in the model sampling away from the subject's true threshold. We did not want to enforce too many priors during data collection and we attempted to prevent such scenarios from occurring by sampling in a fixed grid near (0,0) for initialization trials. However, sensitivity to render camera displacement varied significantly across users, and P10 was more sensitive than other users and in this case the participant only missed a single fixed grid condition. If they had missed a few more trails then the adaptive sampling algorithm would likely have focused on points closer to (0,0).

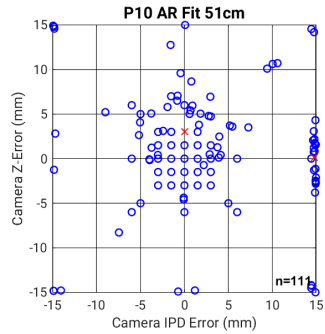

Fig. 13. P10 Lapse trial resulting in poor adaptive sampling.

## 1.4 VR User Study Data

Here we show data for each participant for all VR conditions. In this case we use the same model limit of 80mm for each participant.

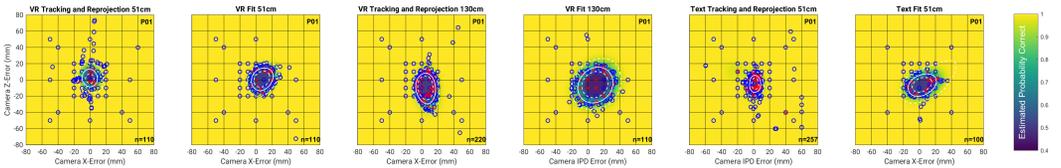

Fig. 14. P01 Raw data for near and far display distances for VR conditions.

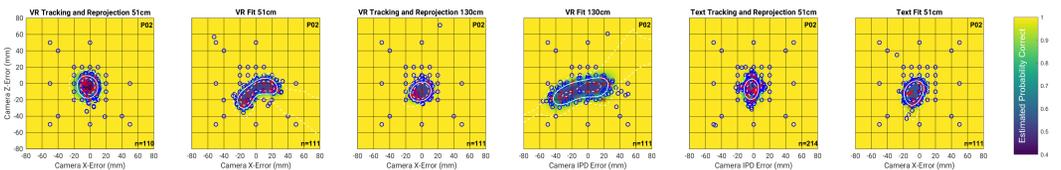

Fig. 15. P02 Raw data for near and far display distances for VR conditions.

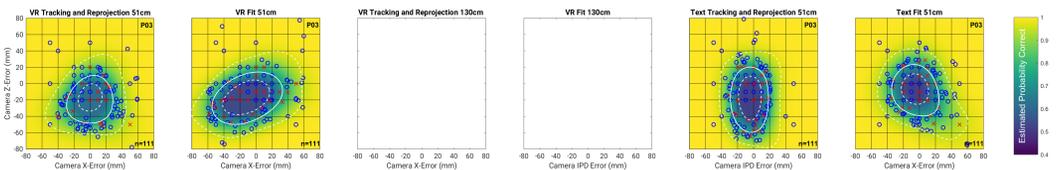

Fig. 16. P03 Raw data for near and far display distances for VR conditions.

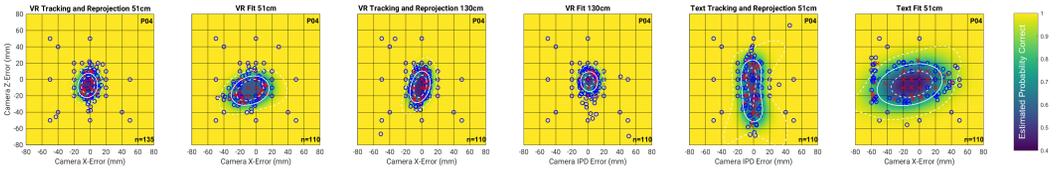

Fig. 17. P04 Raw data for near and far display distances for VR conditions.

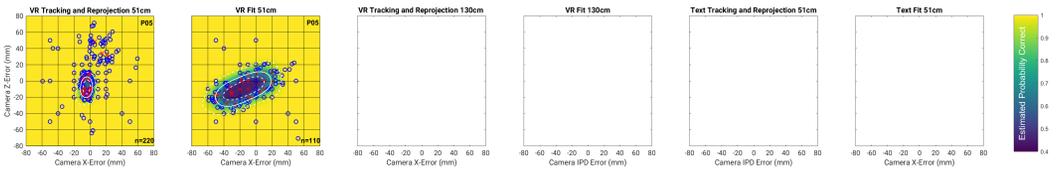

Fig. 18. P05 Raw data for near and far display distances for VR conditions

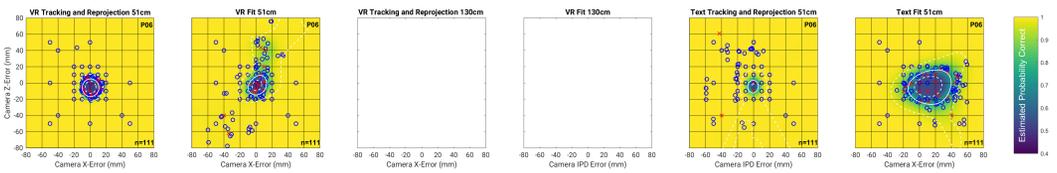

Fig. 19. P06 Raw data for near and far display distances for VR conditions.

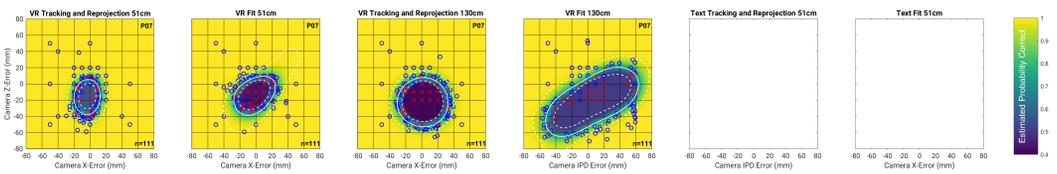

Fig. 20. P07 Raw data for near and far display distances for VR conditions.

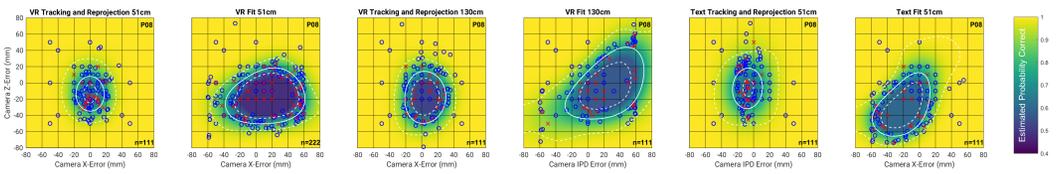

Fig. 21. P08 Raw data for near and far display distances for VR conditions.

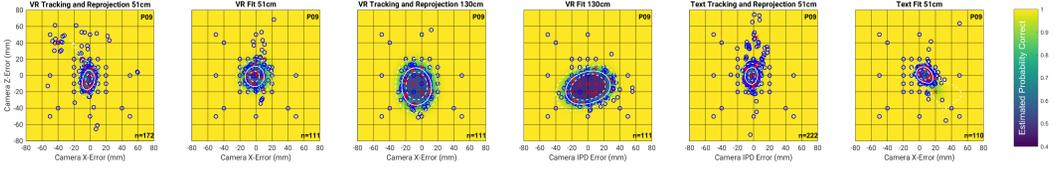

Fig. 22. P09 Raw data for near and far display distances for VR conditions.

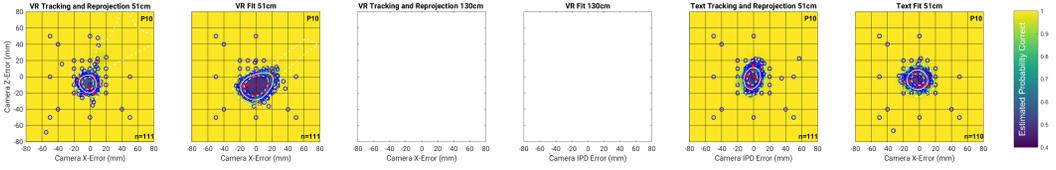

Fig. 23. P10 Raw data for near and far display distances for VR conditions.

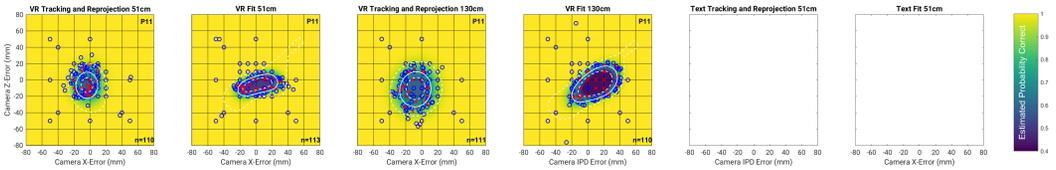

Fig. 24. P11 Raw data for near and far display distances for VR conditions.

## 2 USER STUDY TESTBED

### 2.1 Head Stability on Bite Bar Over Time

It can be uncomfortable to maintain a fixed position on a bite bar over time, and slight amounts of head movement are still possible, especially if users relax their bite. We explored head stabilization in addition to the bite bar to maintain head stability over time by using an adjustable height chin rest to reduce movement between the lower jaw and bite bar. With sufficient pressure from the chin rest, the virtual spheres had zero apparent drift during VOR head and eye movement even when an observer tried to move their heads on the bite bar, however this was uncomfortable and impractical over long periods of time. An alternative was the use of an adjustable forehead rest which was used in combination with the bite bar to stabilize user's heads. In practice, naive observers who were not trained or motivated to stabilize their heads within the user study platform could experience incomplete WLR. However, each of the 11 participants in our user study reported that the nadir of the 5mm spheres remained above the 4mm threaded posts, indicating WLR to within 2.5mm for each subject. We only screened subjects based on visual and stereo acuity, so we believe subjective WLR could be achieved across the general population.

### 2.2 Bite Bar Calibration Hardware

CAD designs are shown for the calibration platform. Note that four beamsplitters (transparent cubes) in the calibration device allow user positioning following the sighting protocol from Hillis and Banks [2001], but were not used in our final bite bar fabrication procedure.

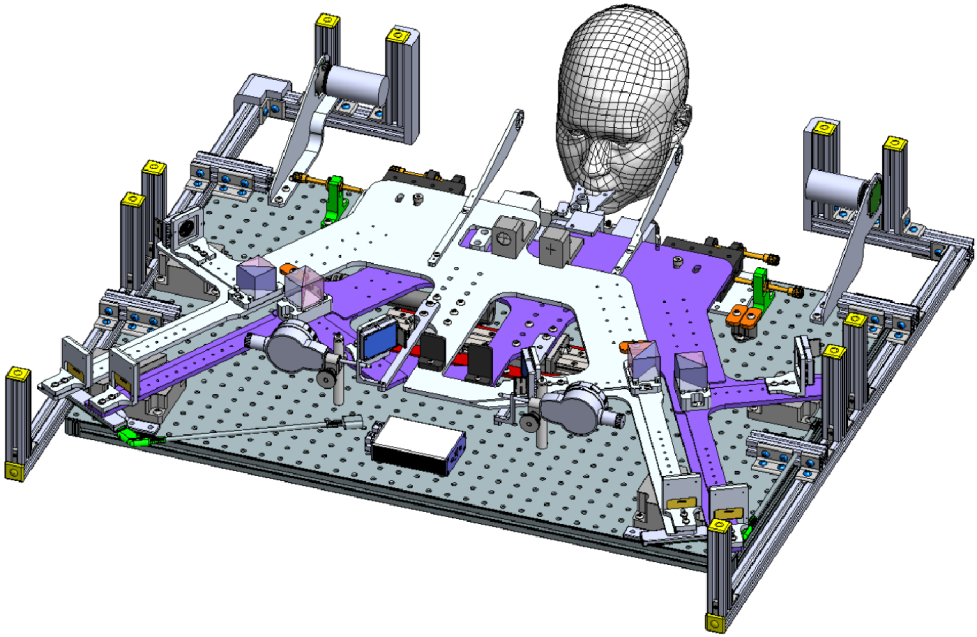

Fig. 25. CAD of Bite bar calibration platform.